\documentclass{JHEP3}
\usepackage{graphicx}

\newcommand{\be}{\begin{equation}}
\newcommand{\ee}{\end{equation}}
\newcommand{\bea}{\begin{eqnarray}}
\newcommand{\eea}{\end{eqnarray}}
\def\IZ{\relax\ifmmode\hbox{Z\kern-.4em Z}\else{Z\kern-.4em Z}\fi}

\newcommand{\IS}{{\bf S}}

\newcommand{\non}{\nonumber \\}\begin{Large}\end{Large}
\def\half{{1 \over 2}} 

\def\pa{{\partial}}

\def\det{\sqrt{-g}}

\def\al{\alpha}

\def\gm{\gamma}

\newcommand{\sbsection}[1]{\vspace{.5cm} \noindent {\it #1}}

 \def\T1{{\bf T}^1}

\def\({\left(} \def\){\right)}
\def\[{\left[} \def\]{\right]}
\def\Schw{~Schwarzschild~}

\preprint{{\tt hep-th/0502034}}

\title{\center{On  Choptuik's  scaling in higher  dimensions}}

\author{ Evgeny Sorkin and Yonatan Oren
  \\
  Racah Institute of Physics\\
  Hebrew University \\
  Jerusalem 91904,
  Israel\\
  {\tt sorkin@phys.huji.ac.il} \  ; \  {\tt yono@phys.huji.ac.il}}

\abstract{We extend Choptuik's scaling phenomenon found in general
  relativistic critical gravitational collapse of a massless scalar
  field to higher dimensions. We find that in the range $4 \leq D \leq
  11$ the behavior is qualitatively similar to that discovered by
  Choptuik. In each dimension we obtain numerically the universal
  numbers associated with the critical collapse: the scaling exponent
  $\gamma$ and the echoing period $\Delta$. The behavior of these
  numbers with increasing dimension seems to indicate that $\gamma$
  reaches a maximum and $\Delta$ a minimum value around $11 \leq D
  \leq 13$.  These results and their relation to the black hole--black
  string system are discussed.}
\begin{document}

\section{Introduction}
\label{sec_intro}

General Relativity (GR) in higher dimensions ($D>4$) has been
receiving increasing attention in recent years. This is largely
motivated by the phenomenological ``brane-world'' and
``large-extra-dimensions'' scenarios that are rooted in String theory,
which is intrinsically higher-dimensional, involving $10$ or $11$
dimensions.  But regardless, GR in itself does not favor the 4D case
especially; rather it is independent of the dimension, and so without
any additional reasons the dimension of space-time may and should be
considered a {\it parameter} of the theory.

Adopting this point of view, one finds that certain properties and solutions of
GR change qualitatively in higher dimensions. In particular, black hole
uniqueness breaks down in $D>4$. This implies the existence of black objects
with non-spherical horizon topologies. Examples include the caged black hole --
black string system in a space-time with compact extra dimensions, see
\cite{barak_review} for a review, and the rotating black hole -- black ring
configurations \cite{MP,blackringER}. The phase transitions between different
horizon topologies seem to lead to a compromise of cosmic censorship in higher
dimensions and to be accompanied by energy outbursts
(see e.g. \cite{kol_explosions}.)
 Moreover, there
exist {\it critical dimensions} above which the qualitative properties of the
spoken phase transition change \cite{TopChange,critdim1,TorusIndication}
\footnote{Another example of
  the qualitative dependence of GR on dimension is the chaotic
  behavior of the spacetime near a spacelike singularity, discovered
  by Belinskii, Khalatnikov \& Lifshitz (BKL). Specifically, above the
  critical dimension $D_{BKL}=10$, the system becomes non-chaotic, see
  the recent review \cite{cosmological_billiards}.}.

In this paper we study the critical collapse of a spherically
symmetric massless scalar field in D-dimensional spacetime, in
search of similar non-trivial dimensional effects. The critical
behavior of black hole formation is a striking example of the
surprising phenomena to be found in gravitational physics. It is
understood that in the dynamical gravitational collapse of matter
fields, initial configurations with very low density will usually
disperse to infinity, while extremely dense clumps will collapse
to form a black hole. What happens in the intermediate case
between these two extremities is much less obvious. In 1992,
Choptuik \cite{Choptuik1} studied this  limiting case numerically,
modeling the collapsing matter as a spherically symmetric
configuration of massless scalar field coupled to gravity. For
each family of initial data (e.g. gaussian, lorentzian etc.),
parameterized by an amplitude $p$, there is a critical amplitude
$p_*$. If $p>p_*$ the final state is a black hole (supercritical
collapse), otherwise the field disperses leaving behind empty flat
space (subcritical collapse). The following discoveries were made
regarding the behavior of this system near the threshold amplitude
$p_*$ in $4D$:

\begin{itemize}

\item In supercritical collapse arbitrarily small black holes are
  formed as $p\rightarrow p_*$. In this limit the black hole mass $M$
  scales as $M \propto (p-p_*)^\gm$, with the pre-factor and $p_*$
  characteristic of the specific family of initial data. The exponent,
  however, is universal and independent of the shape of initial data.
  Its numerical value is $\gm\simeq 0.374$. Later it was also
  discovered that in subcritical collapse the maximal scalar curvature
  encountered before the field disperses scales as $R_{max} \propto
  (p_* -p )^{-2\gamma}$ \cite{GarfinkleDuncan}.

\item The critical solution itself is universal. Namely, following a
  short transient stage, the spacetime converges to a universal
  solution independent of the initial data, remains such for a while
  and then either collapses or disperses, depending on whether $p>p_*$
  or not.

\item The critical solution is discretely self-similar.  Designating
  the critical  solution (collectively for the scalar field and
  the metric)  by $Z_*(r,t)$, this means that
  $Z_*(r,t)= Z_*(r\,e^\Delta,t\,e^\Delta)$, with the empirically
  found $\Delta \simeq 3.44$. The field and metric functions pulsate
  periodically with ever decreasing temporal and spatial scales, until
  a singularity is formed at the accumulation point, $r=t=0$.

\end{itemize}

Similar results were also observed for other kinds of matter and
extensive literature has been written on the subject. For a review see
\cite{GundlachReview} and the references therein \footnote{While a
  majority of authors do not go beyond $4D$, Garfinkle et.  al.
  \cite{Garfinkle6D} have obtained $\gm$ and $\Delta$ in 6D and
  Birukou et. al. \cite{Birukou5D}, got $\gm$ in the critical scalar
  field collapse in $5D$ and $6D$. }.

There are several reasons why we find it interesting to explore the
critical phenomena in higher dimensions: (i) The mysterious universal
numbers that appear in critical collapse, such as the scaling exponent
and the echoing period, are known only empirically. One may wonder if
it is possible to use the {\it dimension as a probe} that can provide
insights or hints into the origin of these numbers; (ii) The Choptuik
critical phenomena, or more precisely its time-symmetric version, is
related to the phase transition in the black-string--black-hole
system\cite{kol_on_choptuik}. It might be possible to learn about the
behavior of that system near the merger point where the topology
changes from that of a black string to that of a black hole by better
understanding the higher-dimensional critical collapse scenario; (iii)
The massless scalar field collapse is probably the simplest model of
dynamical collapse in GR. It is worthwhile to understand what kind of
new features one can expect to find there in higher dimensions.

To realize these goals we designed a numerical code that evolves the space-time and scalar field equations and allows us to study the critical solution in different dimensions.   As is often the case in numerical analysis, we encountered unexpected problems that  demanded special treatment. In addition to the familiar methods used in 4D like ``constrained" evolution and mesh refinement, the code uses analytical series expansion near the origin and ``synthetic friction'' (or ``smoothing") to handle the severe divergences appearing in higher dimensions. We employed this code for $4\leq D\leq 11$.

In all our simulations we find that the near-critical collapse
proceeds in the discretely self-similar (DSS) manner as in the $4D$
case.  For each dimension we obtain the (logarithmic) period of the
pulsations, $\Delta$, and the scaling exponent, $\gamma$, which we
define such that the dimension of length is $[length]\propto
|p-p_*|^\gamma$. These results are summarized in table
\ref{table_del-gam}. In the examined range of dimensions we find that
$\Delta$ decreases and $\gamma$ increases\footnote{There is some
  decrease in $\gamma$ for $D=11$ but we can not definitely determine
  if this reflects a physical behavior or caused by the escalation of
  numerical errors in higher dimensions.  }  with $D$. While we
could not check this explicitly, since our numerical tools did not
take us further than $D=11$, we were tempted to extrapolate our
results beyond that. This {\it extrapolation} seems to indicate that
$\Delta $ reaches a minimum and $\gamma$ has a maximum somewhere in
between $11\leq D\leq 13$, see figures \ref{fig_delta} and
\ref{fig_gamma}. In any case, the dimensional dependence of these
variables is smooth (apparently lacking in divergences).  Moreover,
$\gm$ varies slowly with respect to $D$, and the critical exponent for
mass in D dimensions is $\gm_{\rm mass} = (D-3)\gm$, so as the
dimension grows the mass of a black hole forming above the threshold
increases steeply with $p-p_*$.

In the next section we formulate our problem: we derive the
equations and define the variables. Then in section
\ref{sec_numerics} we describe our numerical scheme.  The results
are presented in section \ref{sec_results}. These and their
relation to the black-hole--black-string system is discussed in
section \ref{sec_discussion}. In the same section we state some
open questions.

\section{Equations and variables}
\label{sec_setup}

We turn now to formulate the physical problem at hand. In
spherical symmetry, we describe the $D$ dimensional asymptotically
flat space-time in double-null coordinates ($u$,$v$)
by the metric:
\be
\label{metric}
ds^2= -\al(u,v)^2 du dv +r(u,v)^2 d\Omega_{D-2}^2,
\ee
where  $ d\Omega_{D-2}^2$ is the metric on a unit $\IS^{D-2}$
sphere and the axis $r=0$ is chosen to be where $u=v$.

The action of the  massless  scalar field $\phi$
minimally coupled to gravity in  $D$ dimensions is given by
\be
\label{action}
S={1\over 16 \pi G} \int R^{} \det d^D x - \int g^{ab}\pa_a\phi \pa_b \phi
\det d^D x.
\ee

The Einstein equations, $R_{ab}=8 \pi G \phi_{,a} \phi_{,b}$ derived
from this action under the metric (\ref{metric}) are
\bea
\label{Einsten_eqs}
r_{,uv} &+& (D-3)\, {r_{,u} r_{,v} + \al^2/4 \over r} = 0, \\
{\al_{,uv} \over \al}  &-&{\al_{,u} \al_{,v} \over \al^2} - {(D-2)(D-3) \over 2}\,  { r_{,u} r_{,v} + \al^2/4 \over r^2}+ 4\pi G \phi_{,u} \phi_{,v} =0, \\
r_{,uu} &-&2 {\al_{,u} \over \al} r_{,u} + {8 \pi G \over D-2}\, r \phi_{,u}^2 =0,  \\
r_{,vv} &-&2 {\al_{,v} \over \al} r_{,v} + {8 \pi G \over D-2}\, r \phi_{,v}^2 =0,
\eea
where the first two are hyperbolic equations of motion, and  the
other two are constraints.

Variation of the action with respect to $\phi$ yields  the scalar field equation of
motion, $\square \phi =0$, which expands to
\begin{equation}
\label{Scalar_eqn}
\phi_{,uv} + {D-2\over 2} \, {\phi_{,u} r_{,v} + \phi_{,v} r_{,u} \over r} =0.
\end{equation}

Following Hamade\&Stewart\cite{HamadeStewart} we formulate the problem as a set
of coupled first order differential equations. Designating  $s\equiv\sqrt{4\pi
G} \phi$ we define
\be \label{definitions}  D1:~ w\equiv s_{,u} ~~~ D2:~ z\equiv s_{,v}
 ~~~ D3:~ f\equiv r_{,u} ~~~ D4:~ g\equiv r_{,v}~~~ D5:~ d\equiv {\al_{,v} \over
 \al}.
\ee
Then the complete set of evolution equations is
\begin{eqnarray*}
\label{Einstein_eqs1}
E1:~~f_{,v} &+& (D-3)\, {f\, g + \al^2/4 \over r} = 0, \\
E2:~~d_{,u} &-& {(D-2)(D-3) \over 2}\,  { f\, g + \al^2/4 \over r^2}+  w\, z =0, \\
C1:~~f_{,u} &-&2 {\al_{,u} \over \al} \, f + {2\over D-2} \, r w^2 =0,  \\
C2:~~g_{,v} &-&2 \,d \,g + {2\over D-2}\, r z^2 =0, \\
S1:~~z_{,u} &+& {D-2\over 2} \, {f\, z +g\, w \over r} = 0,\\
S2:~~w_{,v} &+& {D-2\over 2} \, {f\, z +g\, w \over r} = 0.
 \end{eqnarray*}

Some useful scalar quantities include the Ricci scalar
curvature, which is given by
\be \label{Ricci} R=-{8 \, w\, z \over \al^2} \ee
and the proper time of an observer at the axis,
 \be \label{proptime}
T(u)=\int^u_0 \al(u',u')du'. \ee

The supercritical collapse of a spherical distribution of matter
in $D$ asymptotically flat dimensions results in the formation of
a $D$-dimensional Schwarzschild-Tangherlini black hole, whose
metric reads\cite{Tangherlini}
 \bea \label{SchTang} ds^2 = -f(r) dt^2 &+& f(r)^{-1} dr^2 +r^2
d\Omega_{D-2}^2, \non f(r)&=&1-(r_0/r)^{D-3}, \eea
 The ADM mass of the black hole is $M=(D-2) A_{D-2}
r_0^{D-3}/(16 \pi G)$ in terms of its \Schw radius, $r_0$. $A_{D-2}\equiv 2
\pi^{D/2-1}/\Gamma({D/2-1}) $ is the area of a unit $\IS^{D-2}$ sphere and $G$
is the $D$-dimensional Newton constant.

We define the critical exponent, $\gamma$, such that $|p_*-p|^\gm$ has
dimensions of length. Then it follows \cite{GarfinkleDuncan,Garfinkle6D} that
in $D$-dimensions the maximal curvature (having the dimension of inverse length
squared) achieved in a subcritical collapse scales as $R_{max}\propto
(p_*-p)^{-2\gamma}$ and the mass of the black hole, forming in a supercritical
collapse, behaves as $(p-p_*)^{\gm_{\rm mass}}$ where $\gm_{\rm mass}\equiv
(D-3) \gamma$.

\sbsection{Initial data problem, gauge and boundary conditions}

We specify the initial scalar field profile along an outgoing null
surface $u=u_i=const $, which we choose to be $u_i=0$. To complete
formulation of the initial value problem we note that the choice
of metric (\ref{metric}) is only unique up to the redefinitions
$v'=\chi(v), ~ u'=\xi(u)$. In order to fix this residual gauge
freedom we choose the area coordinate, $r$, along the initial null
surface $u=0$ as $r=v/2$ \footnote{This is chosen to conform with
the conventional definition of characteristic coordinates in flat
space, $v=t+r$ and $u=t-r$.} . In addition, we set $\al=1$ at the
axis on $u_i=v=0$. From here we can obtain all the other functions
on the initial hyper-surface by integration from the origin.

Equations $E1-E2$ and $S1-S2$  are singular at the axis. The
physical solution is, however, perfectly regular there. This
enforces the following boundary conditions on the scalar field and
the metric functions along the axis $r=0$
\bea
\label{bc_axis}
g&=&-f=\half \al, \non
w&=&z, \non
\pa_r s &=&0, \non
\pa_r \al &=&0.
\eea
The source terms which are singular at the axis are then evaluated
by applying l'Hospital's rule.

The actual shape we choose for the initial configuration of the
scalar field is a gaussian shell
\be
\label{s_init}
s(v,u=0) = p\,  \exp\left[- \left( {v-v_c \over
      \sigma}\right)^2 \right]  ,
\ee
where $v_c$ and $\sigma$ are constants and the amplitude $p$ is
the aforementioned strength parameter of the initial data.

\section{The numerical scheme}
\label{sec_numerics}

%
\begin{figure}[t!]
\centering \noindent
\includegraphics[width=11cm]{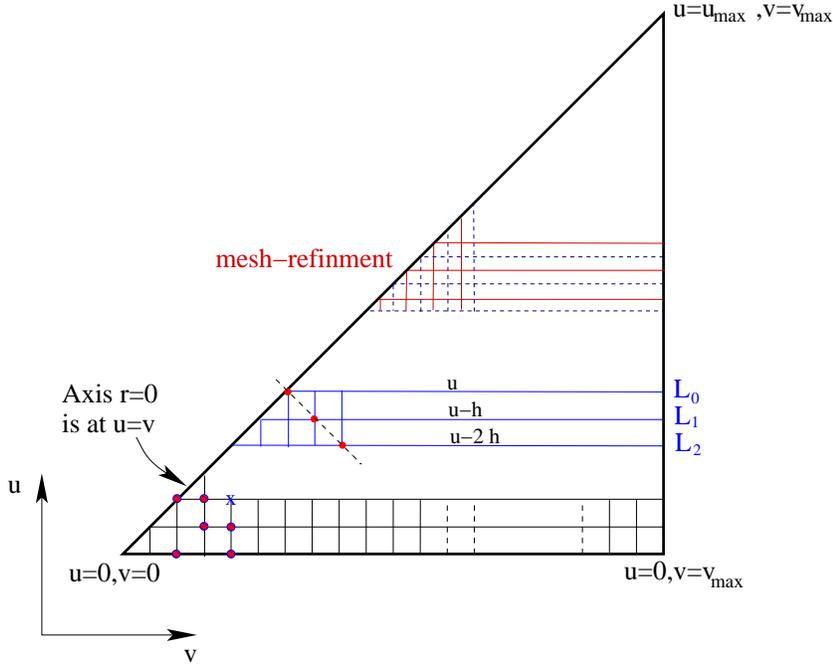}
 \caption[]{ The domain of integration. At any moment, in addition to
   the current outgoing hypersurface, $L_0$ we keep two preceding
   levels $L_1$ and $L_2$. Boundary conditions involving $\pa/\pa r$
   are implemented using 3-point derivatives along the shown diagonal
   line. Mesh refinement is illustrated in the topmost rows.  The
   smoothing of $z$ (or $d$) near the axis at a point marked by a
   cross is done using the values of $z$ (or $d$) at points
   on the past light cone of that point marked with circles.  }
 \label{fig_domain}
\end{figure}
Having laid out the framework, we proceed to describe the
numerical solution of the above equations. We construct a grid in
the $u$-$v$ space as described in figure \ref{fig_domain}. The
primitive computational cell is square, with grid spacings
$h_v=h_u=h$. The initial ingoing wave is specified on an outgoing
hyper-surface of constant $u$, which is stored as a horizontal
line in our grid, see figure\ref{fig_domain} .  Supposing we know
the solution along an outgoing hyper-surface $u-h$ then $d$ and
$z$ are propagated to $u$ using the equations $E2$ and $S1$
correspondingly. Then we integrate the equations $S2, E1, C2, D2,
D4$ and $D5$ from the axis outward along $u$ to get $w, f, g, s,
r$ and $\al$ respectively. Note that we use what is called
``constrained evolution'', namely the constraint C2 is used in the
evolution.  This algorithm is iterated until either the domain of
integration is entirely covered (in the subcritical case) or a
black hole forms (in a supercritical collapse).  The remaining
equations are not used directly (but are used to determine the
boundary conditions) in this scheme, but they must be satisfied by
the solution. We monitor the equations $D3$ and $C1$ and verify
that they are indeed satisfied to a prescribed precision during
the evolution.

We use the 4th order Runge-Kutta (RK4) algorithm to solve all
equations. At every moment in addition to the solution along the
line currently being solved ($L_0$ at $u$) we keep in memory the
solution on the two preceding levels: $L_1$ at $u-h$ and $L_2$ at
$u-2 h$. Without being too memory consuming, this makes the RK4
algorithm more elegantly applicable and also helps us to implement
the axis boundary conditions as sketched in figure
\ref{fig_domain}. The size of the domain is chosen such that the
$v_{max}=u_{max}$ point is located just beyond the accumulation
point.

\sbsection{Series expansion and smoothing  near the axis of evil}

The basic scheme described above works well for collapse in $D=4,5$,
but in higher dimensions the code becomes unstable.  Moreover, for
$D>5$ this scheme is unstable even in {\it flat space}, where only the
wave equations $S1-S2$ are solved (with constant $g=-f=1/2$). This
instability clearly arises near the axis where $r$ is small and the
discretization errors in evaluating the sources are amplified. To
illustrate what happens near the origin let us consider the wave
equation $S2$ in flat space. Say we are solving for $w$ along a
$u=const$ ray at the grid-point, labeled $1$, next to the axis,
labeled $0$. First we rewrite the wave equation as $w_{,v} + [(D-2)/2] w
r_{,v}/r = [(D-2)/4] z/r$, where we used the flat space value $f=-1/2$.
Its formal solution is $w(v_1)=[(D-2)/4]/r^{D/2-1} \int_{v_{0}}^{v_1}
r^{D/2-2} z(v') dv' $.  Using e.g. trapezoidal rule we get
\be \label{Est1} w_1 -w_0 ={D-2\over 8} {h\over r_1} z_{1}  - z_0
+O(h^3), \ee
where we utilized the boundary condition (\ref{bc_axis}), $w_0=z_0$.
Alternatively, approximating the integrand to linear order in $h$ by
Taylor series about $0$ we evaluate the integral analytically to
arrive at
\be \label{Est2} w_1 -w_0= {D-2\over D } \(z_{1} - z_{0}\)
+O(h^2), \ee
Comparing (\ref{Est1}) and (\ref{Est2}) we learn that as $D$ increases
the first term in the r.h.s of (\ref{Est1}) (where $r_{1}=h/2$ ) grows
linearly and dominates while the r.h.s. in the analytical formula
(\ref{Est2}) barely varies. Hence the discrete version (\ref{Est1})
overestimates $w_1$ by a factor\footnote{Methods other than the
  trapezoidal rule can give another factor, which is still
  proportional to $D$, so for large $D$'s the difference between the
  discrete and the analytical formulae increases anyway.}  proportional to
$D$, which clearly causes problems in higher dimensions.

We employ the series expansion method around the axis in equations
$S2$ and $E1$\footnote{The series expansion in E1 also averts error
  amplification in $E2$.}  such that the sources in these equations are
evaluated at the axis.  Then the propagation of $z$ and $d$ in the
$u$-direction is executed as before but with $w$ and $f$ obtained from the
series expansion in all evaluations of sources, required by the Runge-Kutta
algorithm.
We determine the number of points near the origin where the series expansion is
required empirically for every dimension. We found that optimally
this number is approximately $~D$ points. (For example in $D=8$ and for the initial 
grid-spacing of $10^{-3}$ this is done at $9$ grid points.)

Unfortunately it turns out that while the series expansion prolongs
the life-time of the code before crashing, it is not enough to
render it completely stable.  After trying several other methods we
ended up by adding an effective ``friction'' or ``smoothing" in the
$u$-propagated equations ($E2$ and $S1$). Specifically, the value of
the function $z$ (or $d$) at a point near the axis is mixed with
values extrapolated from neighboring points on the past light cone
of that point (see figure \ref{fig_domain}.) Symbolically, the value
of some function $y$  at the point marked by a cross in figure
\ref{fig_domain} is updated according to $y_x=\[y_{e} +3 \omega
\sum_{i=1}^3 y_i \]/(1+3 \omega)$, where $y_e$ is the value obtained
from the evolution equation at the crossed point, and $y_i(u)=2
y(u-h)-y(u-2h)$ are the extrapolated values along the $3$ directions
lying in the past light cone, designated by circles on the figure,
and $\omega$ are the weights. The weights given to the values
obtained from the evolution and from the extrapolation depend on
$D$, but they are essentially comparable. (For the above $D=8$
example the weights $\simeq 1$.) This procedure is  in principle
equivalent to adding a diffusive term in the evolution equations.
The smoothing is applied at the same mesh-points as the series
expansion.

This series-smoothing  symbiosis is a  stable method that
satisfies the constraints $D3$ and $C1$. Its convergence rate,
however, is less than quadratic but never below the linear.  This
is not a serious problem though. The typical run-time of our code
on a PC-class computer with satisfactory precision is on the order
of minutes in spite of this relatively slow convergence. We use
that series-smoothing  method to obtain all the results reported in
this paper.

\sbsection{Mesh refinement}

Close to the threshold amplitude the solution becomes DSS, i.e. it
is self-replicating on decreasing scales.  In order to resolve
this solution the grid spacing must be smaller than the smallest
feature in the solution. A static grid with the spacing necessary
for resolving several echoing periods will require a large amount
of memory and computation time. This is very inefficient, and
moreover unnecessary, because in the early stages the scales
involved are still relatively large, so a very dense grid is
superfluous  to resolve them. This makes dynamical grid refinement
imperative for a realistically feasible scheme. While there exist
completely general adaptive mesh refinement techniques
\cite{BergerOliger}, we use a far simpler method proposed by
Garfinkle \cite{Garfinkle_half}.

The scheme relies on the fact that when $u$ is incremented the grid
point that lies on the axis (at $(v=u)$) enters the region $v<u$ and
effectively leaves the causal past of the domain (see fig.
\ref{fig_domain}). Physically the ingoing ray is reflected from the
origin and becomes an outgoing ray, but grid points which were
assigned to it are lost. Thus the active part of the grid becomes smaller at
large $u$, exactly when more resolution is needed for resolving the
small features of the critical collapse. This unhappy situation can be
prevented by interpolating the remaining points at $v>u$ back into the
original array of points, restoring the original resolution. Following
Garfinkle \cite{Garfinkle_half} we do this only when half of the grid
points cross the axis. Since $v=u$ on the axis, we get a linear
increase in the grid density and a linear decrease in $h$ with $u$.

We use the code to calculate the spacetime near the critical solution in
dimensions $4\leq D \leq11$. For each $D$ we empirically locate the threshold
amplitude $p_*$ by a simple binary search until $p_*$ is found with the desired
accuracy, which is typically one part in $10^{10}-10^{15}$ (the lowest is for
$D=11$). A typical grid has initially between $2000$ and $8000$ grid points in
the initial outgoing surface. After several mesh refinements the step size $h$,
initially of the order of $10^{-3}$ to $10^{-4}$, is reduced by an order of
magnitude or two.

\section{Results}
\label{sec_results}

%
\begin{figure}[t!]
\centering \noindent
\includegraphics[width=10cm]{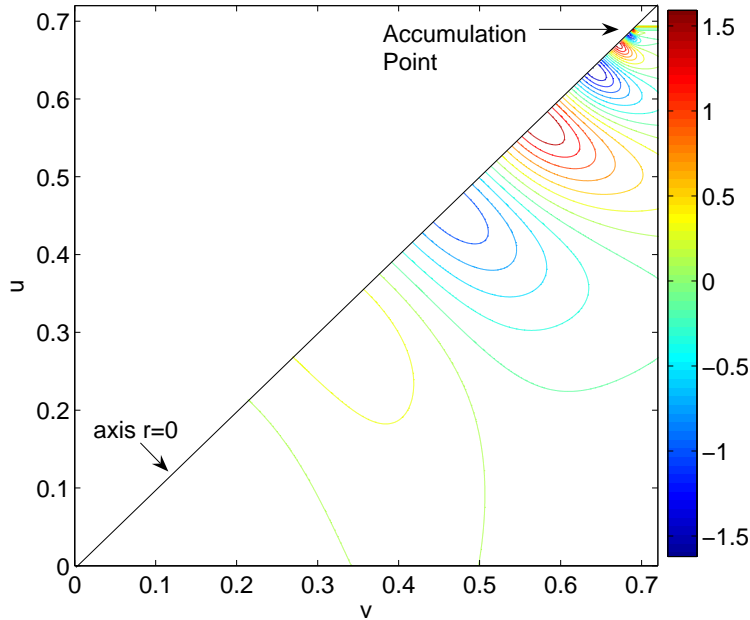}
 \caption[]{ $D=6$: Contours of the  scalar field  profile in
slightly subcritical collapse. After a short transient, the field
oscillates, approaches the accumulation point where the curvature
is maximal and then disperses.
  }
 \label{fig_contours}
\end{figure}
In all dimensions that we examined, the near critical collapse of
the massless scalar field  proceeds in a manner qualitatively very
similar to the familiar $4D$ critical collapse. For $p<p_*$ the
curvature along the axis grows, reaching some maximal value and
then diminishes. In the strong curvature region the scalar field
shows echoing on decreasing scales and subsequently disperses, see
figure \ref{fig_contours}. In the supercritical collapse the field
again rings but the curvature finally diverges and a black hole
forms (the formation of an apparent horizon is signaled by e.g.
the change of sign of $g$, which indicate that outgoing null rays
are tilted back to smaller radii and do not escape to infinity.)
\begin{figure}[t!]
\centering \noindent
\includegraphics[width=10cm]{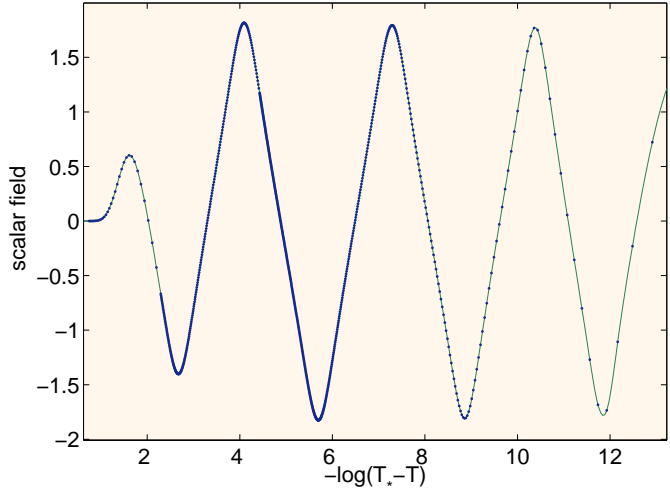}
 \caption[]{ $D=5$: The scalar field  on the
   axis as a function of $\log(T_*-T)$. The period of osculations is
   $\Delta\simeq 3.19$. The actual data is designated by points. The
   distance between the points increases close to $T_*$ indicating the
   decrease of resolution.  }
 \label{fig_s_ax}
\end{figure}

In subcritical collapse we can define the ``accumulation point'' to be
the location of maximal curvature \footnote{For the exactly critical
  solution it coincides with the singularity.}. We also label the
proper time (\ref{proptime}) of an on-axis observer at the
accumulation point by $T_*$. In figure \ref{fig_s_ax} we plot the
scalar field on the axis as a function of $\log(T_*-T)$. After an
initial transient the field becomes periodic in $\log(T_*-T)$,
which is the attribute of DSS behavior in $T$. We were typically
able to observe about $4$ full ringings of the scalar field in
$4D$ and about $3$ for higher $D$. From this data the oscillation
period is computed, and so we get $\Delta(D)$.

\begin{figure}[t!]
\centering
\noindent
\includegraphics[width=10cm]{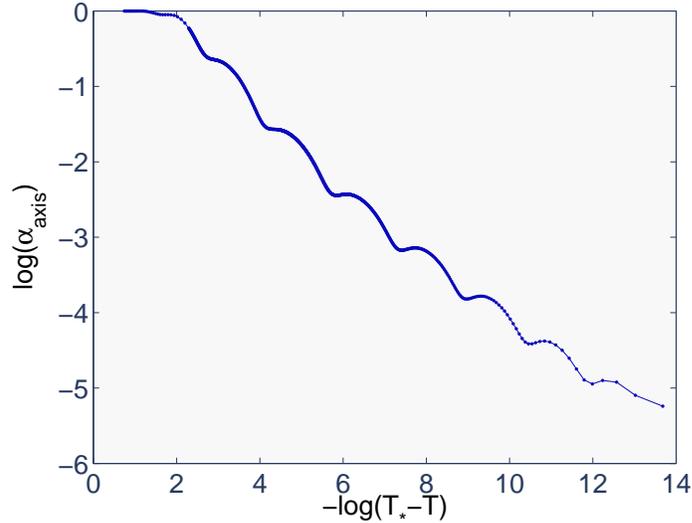}
\caption[]{ $D=5$: The metric  function $\alpha$ on the axis
  decays fast as the accumulation point $T_*$ is approached.  Its
  evolution is accompanied by oscillations whose period ($\simeq 1.6$)
  is half the period of the scalar field ($\simeq 3.19$). }
 \label{fig_a_n3}
\end{figure}
\begin{figure}[h!]
\centering
\noindent
\includegraphics[width=10cm]{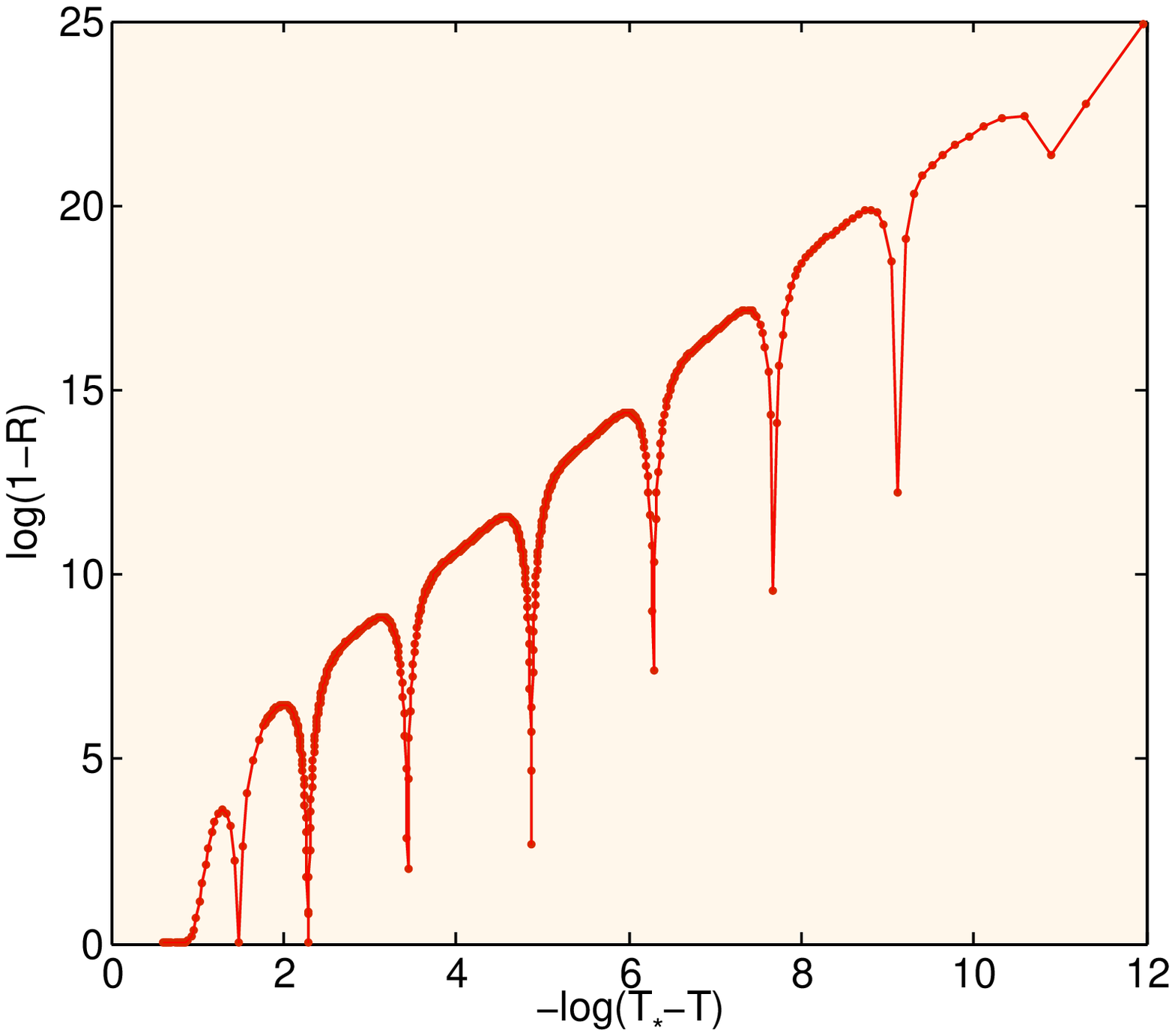}
\caption[]{ $D=7$: Behavior of the curvature as the accumulation
  point $T_*$ is approached. The evolution of the curvature, like
  other metric functions, is accompanied by oscillations.  After each
  pulsation $\log(1-R)$ increases by $\Delta$.  The period of the last
  six echoes is approximately constant and equal to $ \Delta/2 \simeq
  1.41$. Similar behavior (with a different echoing period) is
  observed for other $D$'s as well. }
 \label{fig_riccin5}
\end{figure}

Naturally, not only the scalar field is periodic, but other metric
functions exhibit oscillations as well. In figure \ref{fig_a_n3} we
plot $\al$ at the axis. It shows twice the number of oscillation of
the scalar field, and decreases sharply on approaching the
accumulation point. The evolution of the scalar curvature
(\ref{Ricci}) along the axis is shown in figure \ref{fig_riccin5} to
oscillate with the same period as the scalar field. Since all
variables are periodic, the quantity $\Delta$ can be derived from any
one of them. We found that $\Delta$'s obtained from the various metric
functions are consistent, see table \ref{table_del-gam} for the values
of $\Delta$ in the verified dimensions.

\begin{figure}[h!]
\centering
\noindent
\includegraphics[width=10cm]{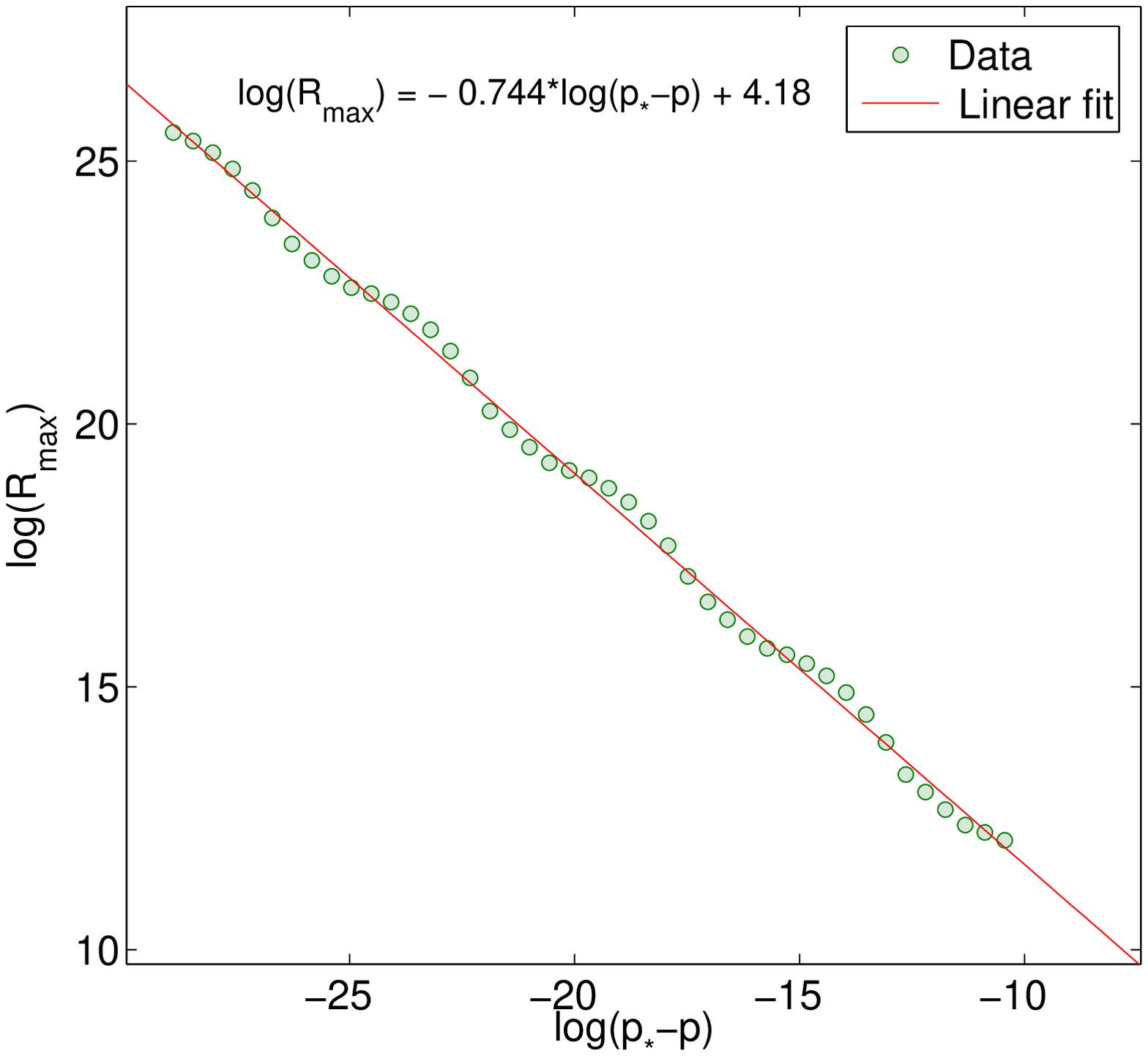}
\caption[]{ $D=4$: The maximal curvature on the axis as a function
of $\log(p_*-p)$. The slope of the linear fit yields $\gm=0.372$
which agrees well with the value cited in the literature ($\simeq
0.374$). }
 \label{fig_gamman2}
\end{figure}

\begin{figure}[h!]
\centering
\noindent
\includegraphics[width=13cm]{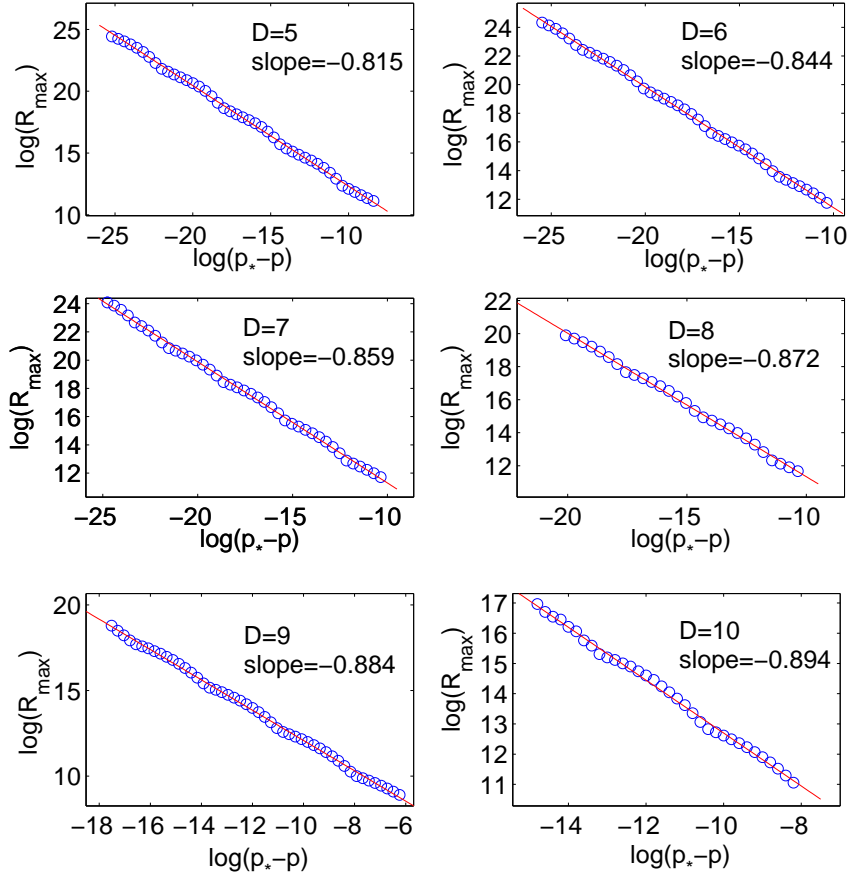}
\caption[]{ The logarithm of the maximal curvature on the axis as a function of
  $\log(p_*-p)$ for different $p$'s in various dimensions.  The slope
  of the linear fit yields $-2 \gm$. Some ``wiggles'', characteristic
  of the DSS solutions, occur about the linear fit. }
 \label{fig_gamman}
\end{figure}
\begin{figure}[t!]
\centering
\noindent
\includegraphics[width=9cm]{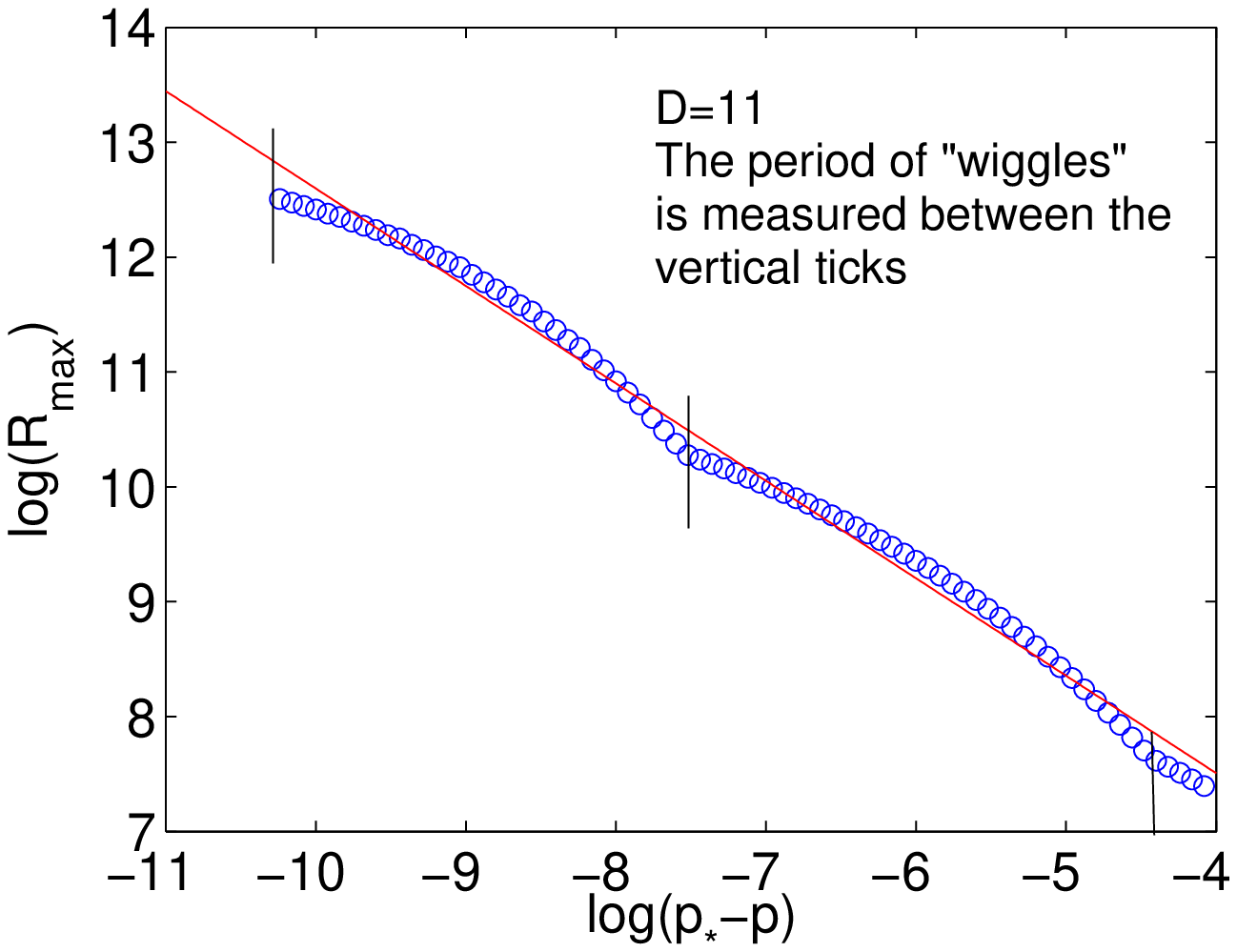}
\caption[]{The logarithm of the maximal curvature on the axis
  as a function of $\log(p_*-p)$ in $11D$.  Because of small number of
  ``wiggles'' and their large  amplitude, the linear fitting approach
  fails to accurately estimate the slope.  In this case we employ  the fact that the
  period of ``wiggles'' ($\simeq 2.85$) is equal to $\Delta/(2\gm)$
  and use this to find $\gamma\simeq 0.44$ from the more accurate $\Delta$. This
  approach is seen in lower dimensions to be consistent with the regular way to obtain $\gamma$.}
 \label{fig_gamman9}
\end{figure}
We evaluated the critical exponent $\gamma$ from the behavior of the
maximal curvature $R_{max}$ in subcritical collapse, and not directly
from the black hole mass scaling in supercritical collapse. The reason
is that it is easier numerically to follow subcritical collapse than
to estimate the black hole mass in supercritical collapse, because of
difficulties in determining the position of the apparent horizon.  (In
addition, as we demonstrate below, it becomes increasingly difficult
to find near-critical black holes in higher dimensions.)

For each dimension we examined the maximal curvature on the axis in
several subcritical simulations. The result for $D=4$ is plotted in
figure \ref{fig_gamman2} and the results for other $D$'s are
summarized in figures \ref{fig_gamman} and \ref{fig_gamman9}. It is
evident from these plots that in all dimensions the maximal curvature
has a dominant power law scaling in $p_*-p$, with the exponent given
by the slope of the linear fit in the figures. The critical exponent,
$ \gamma$, is then the minus half of value of the
slope\footnote{Recall, we define $\gamma$ such that $|p-p_*|^\gamma$
  has dimensions of length.}. $\gm$'s for different dimensions are
listed in table \ref{table_del-gam}.

\begin{figure}[t!]
\centering
\noindent
\includegraphics[width=10cm]{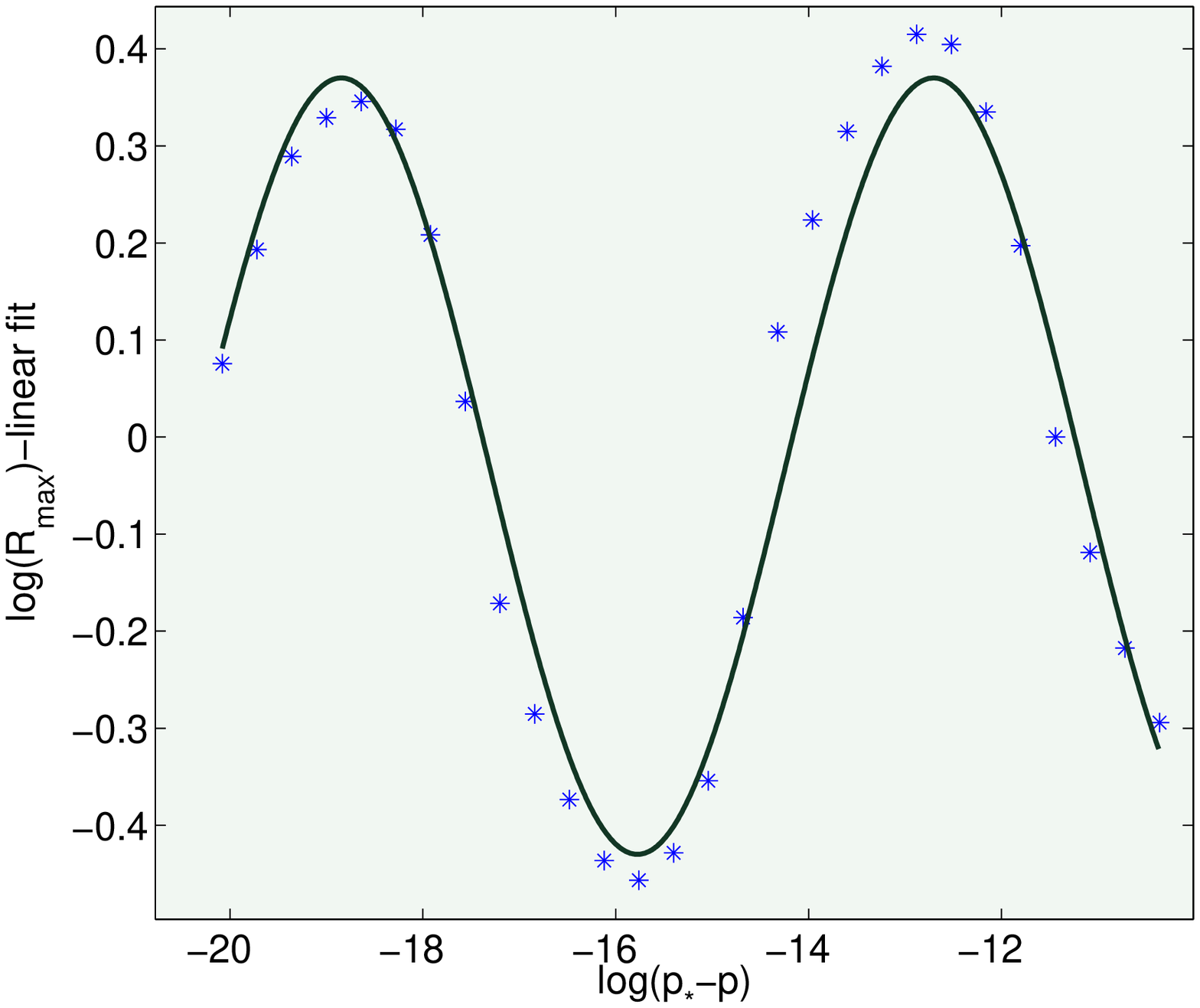}
\caption[]{$D=8$ Normalized plot of  $\log(R_{max}) + 2 \gamma
  \log(p_*-p)$  (stars) and a sine wave (solid
  line). The period   of ``wiggles'' is about $6.18$ and it agrees well with
  theoretical prediction $\Delta/\gamma \simeq 6.22$.  }
 \label{fig_wiggles}
\end{figure}

In addition to the dominant power-law scaling there are some
residual ``wiggles" or fine-structure in the curvature's behavior,
as can be seen in the figures. This was predicted for DSS
solutions and explored in \cite{Gundlach_wiggles,HodPiran_wiggles}
where it was concluded that the ``wiggling" period is
$\Delta/(2\gamma)$. To illustrate this, we subtract from the
measured $R_{max}$ the dominant power-law dependence and plot the
result for $D=8$ in figure \ref{fig_wiggles}. By reflection about
the horizontal axis the data is reorganized to have the doubled
period.  In all dimensions we find the ``wiggling" period to agree
well with the theoretical prediction, $\Delta/\gamma$. This is a
non-trivial test for the numerically computed $\Delta$ and
$\gamma$. Alternatively, since usually we can get $\Delta$ with
somewhat higher precision, $\gamma$ can be computed from the
``wiggles'' period. This is what we do in the $D=11$ case because
in this case the linear fitting approach does not yield a very
accurate result, see figure \ref{fig_gamman9}.

\begin{table}[h!]
\centering
\noindent
\begin{tabular}{|l||l||l|}\hline
  \rm{D}& $~~~~~\Delta $ &$~~~~~~\gamma$\\ \hline
      4 & $3.37 \pm 2\%$            &$0.372 \pm 1\%$ \\ \hline
      5 & $3.19 \pm 2\% $           &$0.408 \pm 2\%$ \\ \hline
      6 & $3.01 \pm 2\%$            &$0.422 \pm 2\%$ \\ \hline
      7 & $2.83 \pm 2\%$              &$0.429 \pm 2\%$ \\ \hline
      8 & $2.70 \pm 3\%$            &$0.436 \pm 2\%$ \\ \hline
      9 & $2.61  \pm 3\%$           &$0.442 \pm 2\%$ \\ \hline
      10 & $2.55 \pm 3\%$            &$0.447 \pm 3\% $ \\ \hline
      11 & $2.51 \pm 3\%$              &$0.44 \pm 3\% $ \\ \hline
\end{tabular}
\caption[]{The echoing-period of the scalar field $\Delta$ and the
scaling exponent $\gamma$ in different  dimensions.
The error in $\Delta$ represents the variation about the mean
$\Delta$ measured on several periods of oscillation. The error in
$\gamma$ originates both from the linear fitting and form the actual
numerical errors in the measured scalar curvature. However it has
been verified by convergence analysis that the dominant error is in
the fitting.The values of $\Delta$ and $\gamma$ that we find for $D=4,6$ agree well with
the numbers found in the literature. }
 \label{table_del-gam}
\end{table}

\section{Discussion}
\label{sec_discussion}

\begin{figure}[t!]
\centering \noindent
\includegraphics[width=10cm]{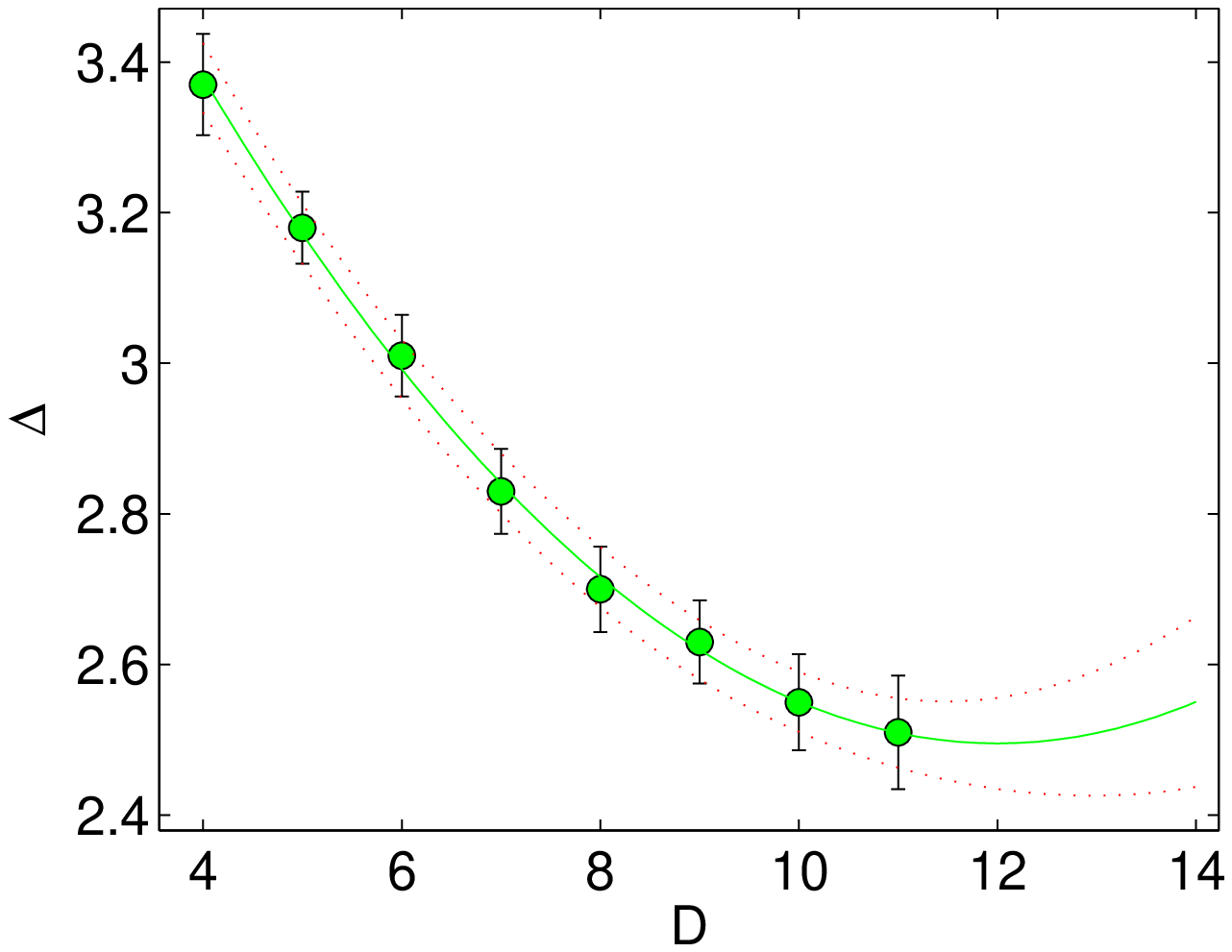}
\caption[]{$\Delta$, including the errorbars, vs the spacetime
  dimension $D$. The solid line designates a $2^{nd}$ degree
  polynomial fitting, the dotted lines show the $95\%$ prediction
  bounds. The speculative extrapolation to higher dimensions suggests
  that $\Delta$ reaches a minimum at $D \simeq 12$.  }
 \label{fig_delta}
\end{figure}

Let us now discuss the dimensional dependence of $\Delta$ and $\gamma$. We find that in the examined range of $D$'s the echoing period $\Delta(D)$ is a decreasing function of dimension, as shown in figure \ref{fig_delta}. The observed dimensional dependence is well behaved in the sense that nothing in this plot forebodes that $\Delta$ will suddenly blow up. Assuming that this regularity continues (otherwise, a discontinuity in $\Delta$ will mark a phase transition) we add, in the same figure, a $2^{nd}$ order polynomial fitting and continue it slightly beyond the last data point.  This {\it extrapolation} indicates that $\Delta$ might have a minimum\footnote{ We do not address the details of this fitting since it has indicative aims only. Yet, one must bear in mind that the robustness of fitting is judged by the allowed variation of the fitting coefficients that still confines the fit within the errorbars.  The proposed (quadratic) fit is significantly better in that sense than all other polynomial fits, making other behaviors than a minimum in $\Delta$ possible but less favored.  } at about $D \simeq 12$.

\begin{figure}[t!]
\centering \noindent
\includegraphics[width=10cm]{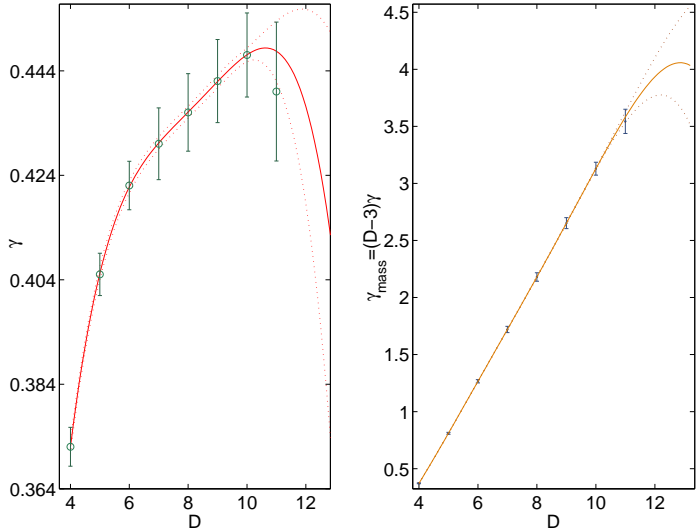}
\caption[]{$\gm$ and $\gm_{\rm mass}$,  with the errorbars, as a
  function of the space-time dimension $D$. The dominant
  dimensional dependence of $\gm_{\rm mass}=(D-3) \gm $ is essentially
  linear increase due to the factor $D-3$, which masks the rather
  subtle variation of $\gm$. The solid line
  designates a 4th order polynomial fit based on the data points in
  the range $4\leq D \leq 10$ (the $\gamma$ value for $D=11$ is {\it
    not} used). The dotted lines show the $95\%$ prediction bounds.
  The extrapolation for larger $D$'s indicates the existence of a
  maximum somewhere in between $10\leq D \leq 12$.  }
 \label{fig_gamma}
\end{figure}

Our computed $\gm(D)$ increases up to $D\leq 10$ but for $D=11$ it
appears to decrease, although the error-bar is consistent with a
constant or even slightly increasing value.
(see figure \ref{fig_gamma}). This apparent decrease in $\gm$ might be
due to a loss of numerical accuracy,
but we are led to suspect the possibility that this is a true physical
effect.  This is suggested by the behavior of $\Delta$. Let us analyze these
two possibilities in more detail.

What could be the numerical cause for an underestimation of $\gamma$?
In our scheme the threshold amplitude $p_*$ can be obtained with
somewhat decreasing accuracy as the dimension grows. Unfortunately our
evaluation of $\gamma$ relies heavily on the value of $p_*$; the
abscissa in figures \ref{fig_gamman2}, \ref{fig_gamman} and
\ref{fig_gamman9} is $-\log(p_*-p)$. Therefore, the quality of the
linear fits in these figures and hence $\gm$, depends acutely on how
well we estimate $p_*$. The effect is such that if the accuracy of
$p_*$ is lowered from 1 part in $10^{14}$ to 1 part in $10^{10}$ a few
percents decrease in $\gm$ is induced. This might well be the reason
we observe a downturn in $\gamma(D)$ at $D=11$.

Having identified the possible cause for a numerical error in
$\gamma$ at $D=11$, we can try to estimate this value independently
of the numerical calculation at this dimension. To this end, we fit
a polynomial (this time of 4th order) to $\gm$ using only data from
$4 \leq D \leq 10$. By extrapolating this fit beyond the last
($D=10$) data point we again observe a change in the trend: $\gm$
seems to reach a maximum at about $D=11$ and then decreases.
Remarkably, the numerically computed $\gm$ value at $D=11$ complies
with this extrapolated behavior. We tend to interpret this  as a
hint that  the observed downturn in $\gamma(D)$ reflects the
physical behavior of the system. This, as well as the behavior of
$\Delta(D)$, marks the appearance
of interesting and non-trivial dimensional dependence associated
with critical collapse. (One might call the dimension where this
occurs a ``critical dimension''.) If so, this intriguing phenomena
clearly deserves better understanding.

While we stress that the allusion to the existence of a special
dimension where the behavior of the the echoing/scaling exponents
changes is based on \textit{extrapolation}, the phenomenon itself is
not entirely unexpected. The Choptuik scaling, albeit in somewhat
different circumstances, is conjectured to appear
\cite{kol_on_choptuik} in the black-string--black-hole system which
does exhibit critical dimensions\footnote{The critical dimensions in
this
  system are: (i) $D_{\rm merger}=10$, above which the local geometry near the
  merger point argued to be  cone-like, and below which this cone-like
  behavior is spontaneously broken\cite{TopChange,KolWiseman}, and (ii) $D_{(\rm 2d ~ order)}=13$ above which the
  phase transition becomes of second order
  \cite{critdim1,TorusIndication}.}. In that system, one considers black objects in a
higher-dimensional spacetime with one compact dimension,
${\mathbb R}^{D-2,1} \times \IS^1$. The known solutions are divided to black
string solutions (whose horizon wraps the compact direction and so has
the $\IS^1\times \IS^{D-3}$ topology), and black hole solutions (with
spherical, $\IS^{D-2}$, horizon topology) localized on the circle.
Consider the space of {\it static} solutions in this system.  The
different phases of solutions (black hole and black strings) are
distinguished by an order parameter in this space, and the transition
between these phases denotes a change in the topology of the solution.
The local analysis \cite{kol_on_choptuik} in the vicinity of the
topology-changing (or merger) point shows that the geometry of the spacetime there
should resemble a time-symmetric version of Choptuik's critical
solution. Formally, the dilaton field $g_{tt}$ plays the role of the
scalar field in Choptuik's case, and the equations are the same. In
both cases the problem is essentially two dimensional.

The similarity between these two systems lends insight into what happens near
the merger point: the geometry here, as speculated in \cite{kol_on_choptuik},
is DSS and it resembles the Choptuik critical solution\footnote{We note,
  however, that it is still not quite clear how to relate the Choptuik solution
  emerging in the collapse situation to its time-symmetric version
  studied in \cite{kol_on_choptuik}.}.  However, we would much prefer
to have information on the dynamic solutions describing this phase
transition.  An attempt to trace the fate of an unstable
string\footnote{It is known that a uniform black string becomes
  unstable \cite{GL1} if ``too thin'' (Actually the relevant parameter
  is the ratio between the \Schw radius of the string and the radius
  of the compact circle).}  in $5D$ was undertaken in
\cite{CLOPPV,GLP}, where it was shown that before the simulation
crashes the black string becomes extremely non-uniform with a very
long and thin neck. The numerical solution in \cite{CLOPPV,GLP} did
not find any fine structure of the critical behavior near the
pinch-off point. We believe that an improved code will discover that
the behavior close to the pinching is similar to what happens in the
axisymmetric near-critical collapse. In the latter case, evidence was
given in \cite{Choptuik_axial} that a non-spherical mode appears
causing bifurcations of the axisymmetric clumps of collapsing matter,
which is reminiscent of the situation expected in the black-string
pinching.

In summary: we have obtained $\Delta(D)$ and $\gm(D)$ for $4 \leq
D\leq 11$ and found clues to the existence of a critical dimension
where the behavior of these functions qualitatively changes.
However, clues are still only that. It is important and
interesting to improve the numerics and discover what really
happens beyond $D=11$. If extrema are present in the functions
$\Delta(D)$ and $\gm(D)$, what are the trends of these functions
for even higher dimensions? Will they be  discontinuous  at  some
dimension, indicating a phase transition? Alternatively, if these
functions are continuous will they tend towards some constant
values? Perhaps $\Delta$ blows up at a certain dimension signaling
that the solution ceases to be DSS and becomes continuously
self-similar (cone-like) beyond that dimension?  We leave these
questions open for future work, that will shed more light on this
stimulating phenomenon.

\vspace{0.5cm} \noindent {\bf Acknowledgments}

We thank Barak Kol for turning our attention to this problem, for
many helpful discussions and for sharing the results of his manuscript
\cite{kol_on_choptuik} prior to publication. ES is supported in part by an ISF grant.


\end{document}